\documentstyle[aps,prl,floats,twocolumn,epsfig]{revtex}
\begin{document}
\title{Control of spin in quantum dots with non-Fermi liquid correlations}
\author{Alessandro Braggio$^{1}$, Maura Sassetti$^{1}$ and Bernhard
  Kramer$^{2}$
  \vspace{1mm}\\
  $^{1}$ Dipartimento di Fisica, INFM, Universit\`{a} di Genova, Via
  Dodecaneso 33, 16146 Genova, Italy \\
  $^{2}$ I. Institut f\"ur Theoretische Physik, Universit\"at Hamburg,
  Jungiusstra\ss{}e 9, 20355 Hamburg, Germany}
\author{{\small (July 18, 2001)} \vspace{3mm}} \author{\parbox{14cm}
  {\parindent4mm \baselineskip11pt {\small Spin effects in the
      transport properties of a quantum dot with spin-charge
      separation are investigated. It is found that the non-linear
      transport spectra are dominated by spin dynamics. Strong spin
      polarization effects are observed in a magnetic field. They can
      be controlled by varying gate and bias voltages. Complete
      polarization is stable against interactions. When polarization
      is not complete it is power-law enhanced by non-Fermi liquid
      effects.}
\vspace{4mm}}} \author{\parbox{14cm} 
 }
\maketitle 

Spin phenomena in transport properties of low-dimensional quantum
systems have become a subject of increasing interest \cite{p98,a98}.
Several fundamental effects have been predicted when controlling
transport of electrons one by one in quantum dots as, for instance,
spin blockade due to selection rules \cite{w95} and parity effects in
the Coulomb blockade \cite{t96,k00}. There are also perspectives of
applications in spin-electronics, quantum computing and communication
\cite{i99}.  Previous works have been focusing on spin transport in
two dimensional (2D) quantum dots connected to non-interacting leads
and in the presence of a magnetic field \cite{r00}, including also an
oscillating magnetic electron spin resonance component \cite{en01}.
Spin transport in circuits with ferromagnetic elements and in the
presence of a Luttinger-liquid interaction \cite{br00,s98,b00} have
been considered.  Of fundamental interest is the spin control of
electron transport in the presence of correlations since in nanoscale
devices the latter are very important.

In the present paper, we derive a general theory for spin and charge
transport through a quantum dot formed in a Luttinger liquid. We
consider spin effects in the presence of a magnetic field.
Specifically, we investigate to what extend non-Fermi liquid behaviour
influences spin polarization. We find that spin-charge separation
strongly affects the current-voltage characteristics. The spin leads
to rich structure in the non-linear differential conductance that
reflects both the collective spin density excitations and the
orientations of the total spin in the quantum dot. A magnetic field in
the quantum dot can spin-polarize the current strongly. This can be
controlled by varying gate- or bias-voltages.  Full spin polarization
can be achieved. Non-complete polarization is power-law enhanced by
the non-Fermi liquid correlations.

We start from a clean Luttinger liquid with spin. The charge
interaction parameter is $g_{\rho}= (1+2V(0)/\pi v_{\rm F})^{-1/2}$
with $v_{\rm F}$ the Fermi velocity and $V(q)$ the Fourier transform
of the electron interaction ($\hbar = 1$). For the exchange
interaction, we assume $g_{\sigma}=1$. The interacting 1D electrons
are mapped, via bosonization, to a harmonic Hamiltonian \cite{v95}.
The low-energy excitations are charge and spin density waves, with
dispersions $\omega_{\nu}(q)=v_{\rm F}|q|/g_{\nu}\equiv v_{\nu}|q|$.
Here, $\nu = \rho, \sigma$ label charge ($\rho$) and spin ($\sigma$).
Spin-charge separation implies $v_{\rho}\neq v_{\sigma}$.  The slowly
varying (on the scale of $2\pi k_{\rm F}^{-1}$) parts of the densities
are given in terms of field operators $\Theta_{\nu}(x)$,
$\nu(x)=\rho_{\uparrow}(x)+p_{\nu}\rho_{\downarrow}(x) \approx
\nu_{0}+\sqrt{2/\pi}\,\partial _{x}\Theta_{\nu}(x)$, with
$p_{\rho}=+$, $p_{\sigma}=-$ and the mean charge and spin densities
$\rho_{0}=2 k_{\rm F}/\pi$ and $\sigma_{0}=0$, respectively.

The quantum dot is formed by barriers $(U_{\rm t}/\rho_{0})
\delta(x-x_{i})$ at
positions $x_{1}<x_{2}$ given by the tunneling Hamiltonian
\begin{equation}
  \label{eq:5}
U_{\rm t}\sum_{\alpha =0,1}\prod_{\nu}
\cos{\frac{\pi}{2}\left(N_{\nu}^{-}+\nu_{0}d
-\alpha\right)}
\cos{\frac{\pi}{2}\left(N_{\nu}^{+}-\alpha\right)}
\end{equation}
with $N_{\nu}^{\pm}=[\Theta_{\nu}(x_{2})
\pm\Theta_{\nu}(x_{1})]\sqrt{2/\pi}$. Physically, $N_{\rho}^{-}$ is
the deviation of the number of electrons from the mean value,
$n_0=d\rho_{0}$, in the interval $d\equiv x_{2}-x_{1}$. The excess
charge then is $Q=-eN_{\rho}^{-}$, and the component of the total spin
of the electrons parallel to the quantization axis (assumed parallel
to the 1D system) $S=N_{\sigma}^{-}/2$. The numbers of imbalanced
electrons and spins between the leads are $N^{+}_{\rho}$ and
$N^{+}_{\sigma}$, respectively. The coupling to the source-drain bias
$V$ and the gate voltage $V_{\rm g}$ is described by $H_{V}=
-e(VN_{\rho}^{+}/2+ V_{\rm g}N_{\rho}^{-}\delta)$, with $\delta$ the
ratio between gate and total capacitances.
The effect of an external magnetic field is described by a local
Zeeman term in the region between the barriers,
$H_{B}=-g_{B}\mu_{B}BN_{\sigma}^{-}/2 \equiv -E_{B}N_{\sigma}^{-}/2$,
with the Land\'e-factor $g_{B}$ \cite{r00,en01}.
 
The currents are calculated as the stationary limits of transferred
charges and spins
$I_{\nu}=I_{\uparrow}+p_{\nu}I_{\downarrow}=(e/2)\lim_{t\to\infty}
\langle\dot{N}_{\nu}^{+}(t)\rangle$. The brackets
include both thermal and statistical averages
over the collective modes at $x\neq x_{1},x_{2}$ with the
density matrix reduced to $N_{\nu}^{\pm}$ \cite{k00}.

For obtaining the non-linear current-voltage characteristics we
consider the dynamics of the system described by the variables
$N_{\nu}^{\pm}$ under the influence of the external fields in the 4D
periodic potential Eq. (\ref{eq:5}). For high barriers, tunneling
between nearest-neighbored minima dominate, with amplitude $\Delta$
that is related to $U_{\rm t}$ via the WKB-approximation
\cite{bra00}. These correspond to processes $N_{\rho}^{-}\to
N_{\rho}^{-}\pm 1$ and $N_{\sigma}^{-}\to N_{\sigma}^{-}\pm 1$
associated with changes of charge and spin numbers in the dot,
respectively, and $N_{\rho}^{+}\to N_{\rho}^{+}\pm 1$ that transfer
current. We consider sequential tunneling with the temperature much smaller
than the dot level spacing.  This can be described by a master
equation for charges and spins with rates,
\begin{equation}
\label{eq:11}
\Xi(E)=\!\!\!\!\sum_{n,m=-\infty}^
{+\infty}\!\!\!\!\!\!W_n^\rho(\epsilon_\rho)W_m^\sigma(\epsilon_\sigma)
\gamma^{(g)}
(E\!-\!n\epsilon_\rho\!-\!m\epsilon_\sigma)\,,
\end{equation} 
where the tunneling rate of a single barrier
\begin{eqnarray}
\label{eq:12}
\gamma^{(g)}(E)=\frac{\Delta^2}{4\omega_{\rm c}}&&
\Big(\frac{\beta\omega_c}{2\pi}\Big)^{1-1/g}\Big|
\Gamma\left(\frac{1}{2g}+\frac{i\beta E}{2\pi}\right)\Big|^2
\nonumber\\
&&\qquad\times\frac{{\rm e}^{-|E|/\omega_c}{\rm e}^{\beta E/2}}
{\Gamma(1/g)}
\end{eqnarray}  
depends on the effective interaction parameter $g/2=g_\rho
g_\sigma/(g_\rho +g_\sigma)$ and the frequency cutoff $\omega_{\rm
  c}$. The weights, $W_n^\nu(\epsilon_{\nu})$, at the discrete
energies $\epsilon_{\nu}$ are ($\beta^{-1} \equiv k_{\rm B}T\ll
\epsilon_{\nu}$)
\begin{eqnarray}
W_n^\nu(\epsilon_{\nu})\approx \left(\frac{\epsilon_{\nu}}
{\omega_c}\right)^{1/2g_\nu}\frac{\Gamma(1/2g_\nu+n)}
{n!\,\Gamma(1/2g_\nu)}\,{\rm e}^{-\epsilon_{\nu} n/\omega_c}\theta(n)\,.
\label{eq:13}
\end{eqnarray}

In order to understand the rather complex behaviour of the transport
spectra it is useful to recall the characteristic energy scales in
(\ref{eq:11}). These are the {\em discretization energies}
corresponding to charge and spin modes in the quantum dot relative to
the energy of the ground state,
\begin{equation}
  \label{eq:14}
  \epsilon_{\nu} \equiv\omega_{\nu}(q=\pi/d) = 2 g_{\nu}E_{\nu}\,,
\end{equation}
with the {\em addition energies} for charge and spin
\begin{equation}
  \label{eq:15}
E_{\nu}=\pi v_{\rm F}/2g_{\nu}^{2}d\,.
\end{equation}
Without interaction, the addition energies are
$E_{\rho}=E_{\sigma}=\pi v_{\rm F}/2d=E_{\rm P}\neq 0$, due to the
Pauli principle, and the discreteness of the dot levels.  On the other
hand, for strong Coulomb interaction, $E_{\rho}\propto V(q\to 0)\gg
E_{\sigma}$ \cite{k00}.

The addition energies determine the ground state energy of $n$ charges
with the half-integer total spin $S\equiv s_{n}/2$,
$E_{0}(n)=[E_{\rho}\left(n-n_{\rm
    g}\right)^{2}+E_{\sigma}s_n^{2}-E_{B}s_{n}]/2$.  The reference
particle number $n_{\rm g}\equiv eV_{\rm g}\delta/E_{\rho}+n_0$ is
defined by the gate voltage. The energy differences of the many-body
states of $n+1$ and $n$ electrons are
\begin{eqnarray}
  \label{eq:15}
  \mu(n,s,l,m) &=& \frac{E_{\rho}}{2}\left(2n+1-2n_{\rm g}\right)
+\frac{E_{\sigma}}{2}(s_{n+1}^{2}-s_n^{2})\nonumber\\
&-&\frac{E_{B}}{2}(s_{n+1}-s_{n})+
l\epsilon_{\rho}+m\epsilon_{\sigma}\,.
\end{eqnarray}
Positive or negative integers $l$ and $m$ denote the differences of
the numbers of charge and spin excitation quanta with the energies
(\ref{eq:14}), respectively. These do not change neither the number of
particles nor the total spin in the quantum dot. The energy
differences $\mu(n,s,l,m)$ play the role of chemical potentials of the
dot and define the transport regions. For symmetric bias, for
instance, the condition $V/2>\mu(n,s,l,m)>-V/2 $ defines the allowed
transport channels. For $V\to 0$ and $E_{B}\ll E_{\sigma}$ one finds
the Coulomb blockade peaks at gate voltages $V_{\rm g}$ that
correspond to $\mu(n,s,l,m)$. However, due to the spin, the separation
of the peaks depend on the parity of $n$, $\Delta V_{\rm g}^{n+1,
  n}=[E_{\rho}+(-1)^{n+1}(E_{\sigma}- E_{B})]/e\delta$. Without
interaction, one has to replace here $E_{\rho}=E_{\sigma}\equiv E_{\rm
  P}$, in order to get the separation of the linear conductance peaks.

\begin{figure}
\setlength{\unitlength}{1cm}
\begin{picture}(7.0,4.2)(0,-0.4)
\put(-0.8,-20.7){\epsfxsize=16.5cm\epsfysize=26.0 cm 
              \epsffile{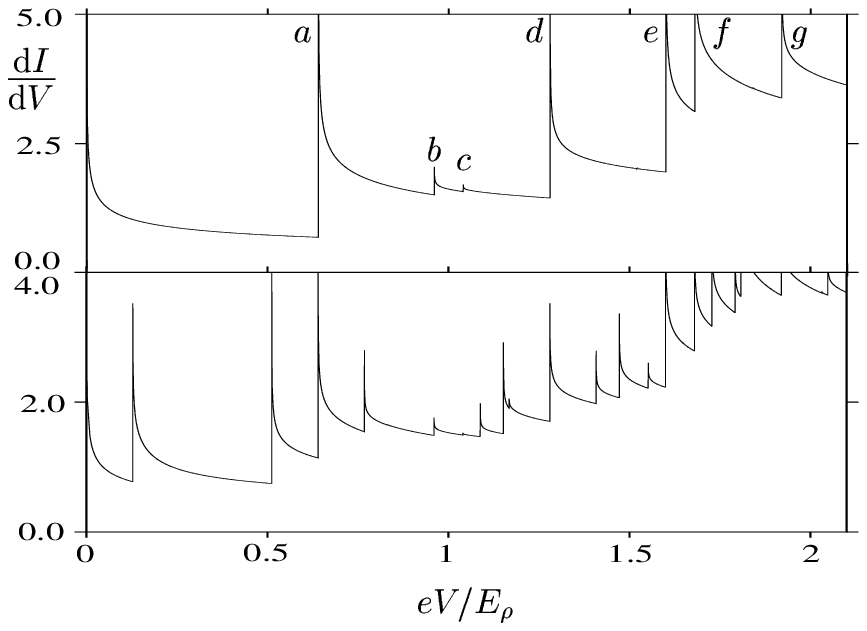}}
\end{picture}\par
\vskip1.35cm
\caption{Differential conductance ${\rm d}I/{\rm d}V$ as a 
  function of the source-drain bias $eV/E_{\rho}$ for $g_{\rho}=0.4$
  ($\omega_{\rm c}=10^{5}E_{\sigma}$, units $10^{-3} (\omega_{\rm
    c}/E_{\sigma})^{(g-1)/2g} e^2\Delta^2/4\omega_{\rm c}^2$). Top:
  $E_{B}=0$, $n_{\rm g}=0.58$; bottom: $E_B=0.4\,E_{\sigma}$, $n_{\rm
    g}=0.548$.}
  \label{braggiofig:1}
\end{figure}

In the following, we consider the limit $T=0$. Results for the
differential conductances ${\rm d}I/{\rm d}V$ for $E_{B}=0$ and
$E_{B}=0.4\,E_{\sigma}$ as functions of $V$ are shown in
Fig.~\ref{braggiofig:1}. Zero bias voltage has been assumed at the
position of a conductance peak corresponding to an
$n$(even)-to-$(n+1)$ ground-state-to-ground-state transition. The
differential conductance shows sharp peaks at bias voltages $V_{nlsm}$
at which a new transport channel enters the above bias voltage window.
Above $V_{nlsm}$, the conductance as a function of $V$ drops according
to the interaction-induced non-Fermi liquid power law
$(V-V_{nlsm})^{1/g-2}$.

From (\ref{eq:15}) one can easily identify the spectral origins of the
peaks in the conductance spectra. At low voltages, the spectra are
completely dominated by spin excitations due to spin-charge
separation. The discretization and addition energies corresponding to
the spin are factors $g_{\rho}$ and $g_{\rho}^{2}$, respectively,
smaller than those corresponding to the charge. For $B=0$ peak ($e$)
corresponds to a charge density excitation at $\epsilon_{\rho}$, while
($f$) is due to the ground-state to ground-state transition at
$E_{\rho}-E_{\sigma}$. All of the other peaks in
Fig.~\ref{braggiofig:1} are spin-related. Because $g_{\sigma}=1$ the
transition at $2E_{\sigma}$ is degenerate with the spin density
excitation at $\epsilon_{\sigma}$ (peak ($a$), with multiples ($d$)
and ($g$)).  A finite exchange would remove this degeneracy and
discriminate between spin addition energies and spin density waves.
The two small features ($b$) and ($c$) are combinations of the
excitations ($e$) and ($f$) with $E_{\sigma}$, they corresponds to
$\epsilon_{\rho}-2E_{\sigma}$ and $E_{\rho}-3E_{\sigma}$,
respectively. In the non-interacting limit, the peaks in the {\em
  non-linear} differential conductance appear at bias voltages that
are multiples of $4E_{\rm P}$ due to the absence of spin charge
separation. Apart from the peak at $V=0$, all of the peaks are absent
without electron interaction, for the bias voltages in
Fig.~\ref{braggiofig:1} (top).

The spin-related features are even more strikingly displayed in the
spectra for $E_{B}=0.4\,E_{\sigma}$ (Fig.~\ref{braggiofig:1} bottom). All
of the peaks in Fig.~\ref{braggiofig:1} (top) acquire Zeeman side bands
corresponding to energies $E_{\rm peak}\pm E_{B}$. Exceptions are
($b$) and ($c$). They only have sidebands $E_{\rm peak}+E_{B}$ since
the initial states corresponding to the lower sidebands cannot be
occupied by electrons entering the dot at $T=0$.

\begin{figure}
\setlength{\unitlength}{1cm}
\begin{picture}(7.0,4)(0,-0.4)
\put(-1.6,-23.5){\epsfxsize=22cm\epsfysize=30.0 cm 
              \epsffile{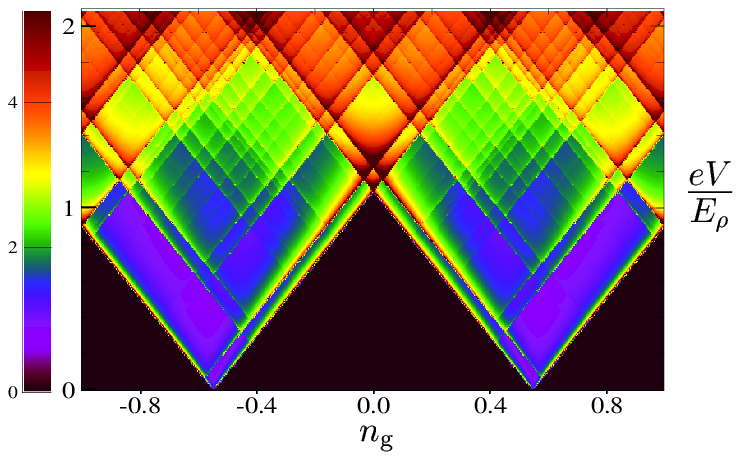}}
\end{picture}\\\par
\vskip0.4cm
\caption{Density plot of the differential conductance 
  in the ($eV/E_{\rho}$,$n_{\rm g}$)-plane for interaction
  $g_{\rho}=0.4$ and magnetic field $E_B=0.4\,E_{\sigma}$
  ($\omega_{\rm c}=10^{5}E_{\sigma}$, color code (left) with units
  $10^{-3} (\omega_{\rm c}/E_{\sigma})^{(g-1)/2g}
  e^2\Delta^2/4\omega_{\rm c}^2$).}
\label{braggiofig:2}
\end{figure}

As an example of the rich spin-related structure in the transport
spectrum a density plot of the differential conductance in the
($eV/E_{\rho}$,$n_{\rm g}$)-plane for $E_B=0.4\,E_\sigma$ is shown in
Fig.~\ref{braggiofig:2}. The black regions correspond to the Coulomb
blockade. When increasing the bias voltage the the excitation spectrum
displays a considerable number of spin-related transitions. With
non-vanishing exchange interaction, $g_{\sigma}\neq 1$, accidental
degeneracies due to spin addition and excitation are lifted and the
spectrum becomes even more complex.

\begin{figure}
\setlength{\unitlength}{1cm}
\begin{picture}(9.0,4)(0,-0.4)
\put(-2.4,-19.7){\epsfxsize=23cm\epsfysize=27 cm 
              \epsffile{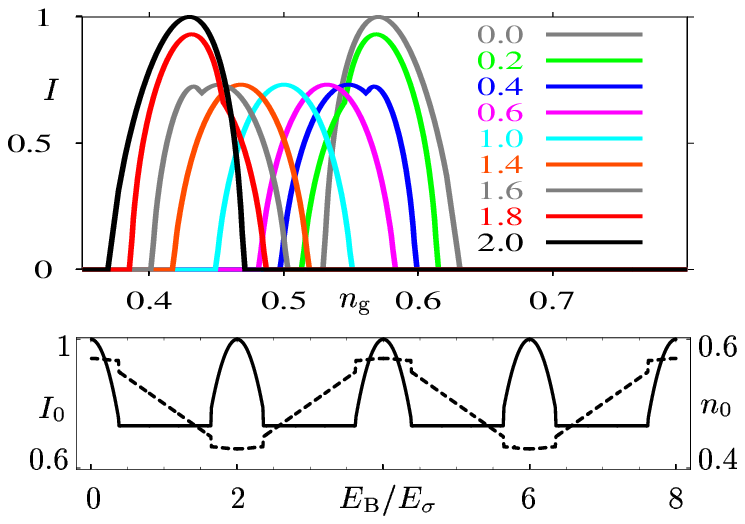}}
\end{picture}\\\par
\vskip0.4cm
\caption{Top: Current $I$, normalized to the value at $B=0$, 
  as a function of the gate voltage $n_{\rm g}$ for different magnetic
  fields $E_{B}/E_{\sigma}$ (inset) for $g_{\rho}=0.4$ and at $eV=
  0.1\,E_{\rho}$ ($\omega_{\rm c}=10^{5}E_{\sigma}$). Bottom: position
  of maximum of peak, $n_{\rm 0}$ (dashed, right scale) and current at
  peak maximum, $I_{\rm 0}$ (full line, left scale) as a function of
  $E_{B}/E_{\sigma}$.}
\label{braggiofig:3}
\end{figure}

Figure \ref{braggiofig:3} shows the behaviour of a current peak for fixed
bias, $I(n_{\rm g})$, when changing the magnetic field. As $B$ is
changed, peak height and position vary with periods $2E_{\sigma}$ and
$4E_{\sigma}$, respectively (Fig.~\ref{braggiofig:3} bottom). This can be
understood by considering the processes that contribute towards the
current. We start by discussing the peak position for $V\to 0$. For
small $B$ and keeping $n_{\rm g}$ as to match the maximum of the peak
(Fig.~\ref{braggiofig:3} top) one finds from (\ref{eq:15}) that with
increasing $B$ one has to adjust $n_{\rm g}$ to lower values $\propto
- E_{B}/2$ since $s_{n+1}-s_{n}=+1$ which corresponds to the
$s_{n}=0\to s_{n+1}=1$ transition. When $E_{B}\ge 2E_{\sigma}$, the
energy of the state $s_{n}=2$ becomes lower than that of the $s_{n}=0$
state such that transport gets now support from transitions
$s_{n}=2\to s_{n+1}=1$ with $s_{n+1}-s_{n}=-1$ while $s_{n+1}+s_{n}=3$
instead of $1$. When increasing $B$ further, $n_{\rm g}$ has now to
be adjusted to higher values $\propto +E_{B}/2$ in order to compensate
for the Zeeman shift. The original position is reached after a total
change of $B$ corresponding to $E_{B}=4E_{\sigma}$. For $V\neq 0$, the
peak position remains unchanged as long as energies of the transitions
are inside the interval $(-V/2,V/2)$.

\begin{figure}
\setlength{\unitlength}{1cm}
\begin{picture}(8,4.5)(0,-0.4)
\put(-1.03,-18.7){\epsfxsize=20cm\epsfysize=25.0 cm 
              \epsffile{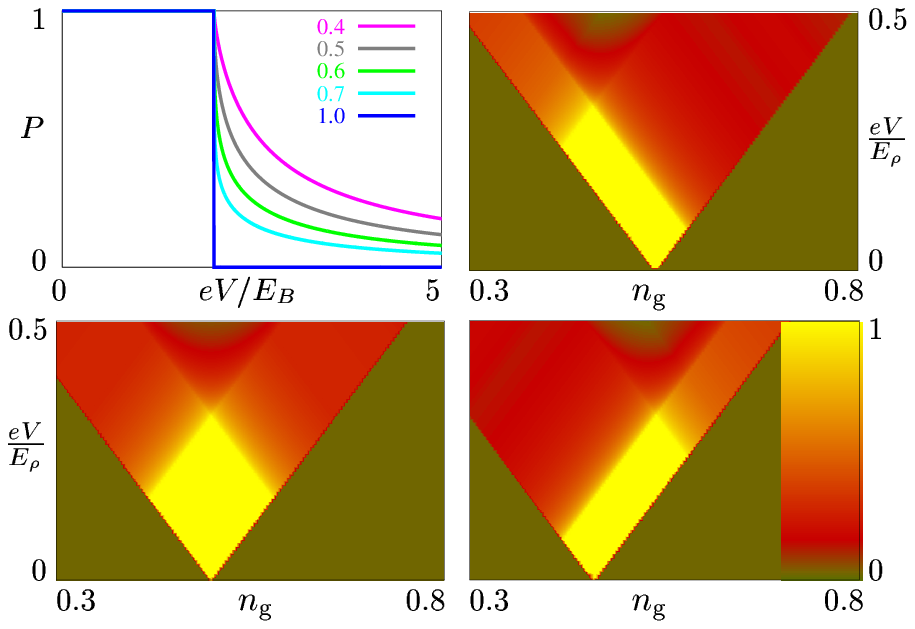}}
\end{picture}\\\par
\vskip0.5cm
\caption{Current polarization $P=I_{\sigma}/I_{\rho}$
  for different interactions $g_{\rho}$ (inset) as a function
  of $eV/E_{B}$ for $E_{B}=0.4\,E_{\sigma}$ and $n_{\rm g}$ at the
  maximum of the conductance peak (top left); $P$ in the plane
  $(eV/E_{\rho},n_{\rm g})$ for $g_{\rho}=0.4$ and
  magnetic field $E_{B}=0.5\,E_{\sigma}$ (top right),
  $E_{B}=E_{\sigma}$ (bottom left) and $E_{B}=1.5\,E_{\sigma}$
  (bottom right, includes color code).}
\label{braggiofig:4}
\end{figure}

For understanding the peak height we consider $V\neq 0$. The bias $V$
gives the width of the current peak. First, we remember that for
$E_{B}\ll eV$, the current is due to two transitions, namely
$s_{n}=0\to s_{n+1}=\pm 1$. This leads to the asymmetry of the
(non-spin polarized) peak for $E_{B}=0$ that can be observed in
Fig.~\ref{braggiofig:3} (top). When $E_{B}\approx eV$ the contribution of
the ground-to-excited-state transition $s_{n}=0\to s_{n+1}=-1$ is
suppressed. Further increasing $B$, the current peak becomes
symmetric, completely polarized (cf. Fig. \ref{braggiofig:4}), its height
is reduced and remains constant. For $E_{\sigma}\ll E_{B} <
2E_{\sigma}$ transport gets support also from transition $s_{n}=2\to
s_{n+1}=1$ such that the peak height starts to become again asymmetric
and to increase until, exactly at $E_{B}=2E_{\sigma}$, both
contributions are equally important. Then, the current peak acquires
the same shape as for $E_{B}=0$, but reflected at $n_{\rm g}=1/2$.
Increasing $B$ produces oscillatory behavior due to further changes of
the values of $s_{n}$ and $s_{n+1}$. The current has its maximum
value at $n_{\rm g}=1/2$ whenever $E_{B}$ is an odd multiple of
$E_{\sigma}$. This fully polarized current states are also reflected
in the digital behaviour \cite{cio00} displayed in
Fig.~\ref{braggiofig:3} (bottom). The latter becomes less stable when
increasing the bias voltage $V$.

Figure~\ref{braggiofig:4} shows the behaviour of the current
polarization, $P=I_{\sigma}/I_{\rho}$. The top-left panel shows $P$
as a function the source-drain voltage for a given magnetic field and
different interaction strengths, with $n_{\rm g}$ at the maximum of
the linear conductance. The polarization is complete for $eV\le 2E_B$,
independently of the interaction. When $eV>2E_B$, $P$ decreases as a
function of $V$ according to a non-Fermi liquid power law and is
higher for stronger interaction. Thus, correlations {\em enhance} the
polarization when the latter is not complete.

The other three panels of Fig.~\ref{braggiofig:4} show density-plots of
the polarization for fixed interaction. Details of the behaviour can
be understood from the discussion related to Fig.~\ref{braggiofig:3}.
Varying the magnetic field, complete polarization, $P=1$ (yellow), is
transferred between different regions of the $(eV,n_{\rm g})$-plane.
The top-right panel corresponds to $E_B= 0.5\,E_{\sigma}$ were the
dominating $s_{n}=0\to s_{n+1}=1$ channel leads to complete spin
polarization near the left-hand edge of the region of non-zero
current. Near $E_{B}=E_{\sigma}$ (bottom-left) the current peak is
symmetrically spin polarized (for small $V$, yellow diamond).
Increasing the magnetic field fully polarizes the right-hand edge of
the current region ($E_B=1.5\,E_{\sigma}$, bottom-right). Exactly at
$E_{B}=2E_{\sigma}$ the two involved transport channels have equal
weight and spin polarization is exactly zero, as for $E_{B}=0$. By
further increasing $B$, the behaviour of the polarization is reversed,
$P=-1$. It can be displayed by the same panels, but in opposite
direction. The periodicity of $P$ with respect to $E_B$ corresponds to
$4E_{\sigma}$.

In conclusion, we have shown how one can control the spin properties
of the transport through a 1D quantum dot embedded in a non-Fermi
liquid by changing magnetic field, and gate- and source-drain
voltages. This is due to the interaction that separates the energy
scales of the charge and the spin excitations such that $E_{\sigma}\ll
E_{\rho}$. Complete spin polarization can be achieved in spite of the
presence of correlations. Once it is achieved, it is not influenced by
the interaction. When polarization is not complete, it is enhanced by
the non-Fermi liquid correlations. This shows that the electron spin
is crucial for understanding non-linear transport in 1D quantum dots.

The above results have been obtained for $T=0$.  When $T>0$, we expect
temperature-induced Luttinger liquid power-law broadening of the
conductance peaks, and correspondingly a smearing of the spin
polarization features which is the subject of future work.  At very
low temperatures, one would expect that coherent tunneling processes
dominate. Including spin, these lead to the well-known quantum-dot
Kondo physics which is not considered in our approach \cite{fu98}.
Therefore, the above results apply to temperatures higher than the
Kondo temperature.
 
Our results for the transport spectra are consistent with several of
the non-linear features observed recently in a one dimensional (1D)
quantum dot formed by two impurities in a cleaved-edge overgrowth
quantum wire \cite{a00}. More detailed experiments, however, are
needed in order to test the above predictions, especially concerning
the control of the spin. We expect that the effects can be observed in
the transport through double barriers formed in
cleaved-edge-overgrowth quantum wires, and in carbon nanotubes
\cite{dek01}.

Acknowledgements: This work has been supported by Italian MURST
via PRIN 2000, by the EU within TMR and RTN programmes, and by the
DFG.

\end{document}